\documentclass[journal=jpclcd,manuscript=letter]{achemso}
\usepackage{amssymb,amsmath,cmbright}
\usepackage{lmodern}
\usepackage[utf8]{inputenc}
\usepackage{chemformula} 
\usepackage{chemfig}
\usepackage[T1]{fontenc} 
\usepackage{graphicx}
\usepackage{dcolumn}
\usepackage{multicol}
\usepackage{multirow}
\usepackage{threeparttable}
\usepackage[table]{xcolor}
\usepackage{siunitx}
\DeclareSIUnit\angstrom{\text{Å}}
\usepackage{csquotes}
\usepackage{hyperref}
\hypersetup{
    colorlinks=true,
    linkcolor=cyan,
    filecolor=magenta,      
    urlcolor=cyan,
    citecolor=cyan,
    }

\DeclareMathVersion{sans}
\SetSymbolFont{operators}{sans}{OT1}{cmbr}{m}{n}
\SetSymbolFont{letters}{sans}{OML}{cmbr}{m}{it}
\SetSymbolFont{symbols}{sans}{OMS}{cmbr}{m}{n}
\SetMathAlphabet{\mathit}{sans}{OT1}{cmbr}{m}{sl}
\SetMathAlphabet{\mathbf}{sans}{OT1}{cmbr}{bx}{n}
\SetMathAlphabet{\mathtt}{sans}{OT1}{cmtl}{m}{n}
\SetSymbolFont{largesymbols}{sans}{OMX}{cmex}{m}{n} 

\mathversion{sans}

\SectionNumbersOn

\title{Limitations of MRSF-TDDFT for Applications in Photochemistry}

\author{Ji\v{r}\'{i} Jano\v{s}}
\email{jiri.janos@vscht.cz}
\affiliation[VSCHT]
{Department of Physical Chemistry, University of Chemistry and Technology, Technická 5, Prague 6, 166 28, Czech Republic}
\alsoaffiliation[UB]
{School of Chemistry, University of Bristol, Bristol BS8 1TS, United Kingdom}

\author{Andrew J. Orr-Ewing}
\affiliation[UB]
{School of Chemistry, University of Bristol, Bristol BS8 1TS, United Kingdom}

\author{Basile F. E. Curchod}
\affiliation[UB]
{School of Chemistry, University of Bristol, Bristol BS8 1TS, United Kingdom}

\author{Petr Slav\'{i}\v{c}ek}
\email{petr.slavicek@vscht.cz}
\affiliation[VSCHT]
{Department of Physical Chemistry, University of Chemistry and Technology, Technická 5, Prague 6, 166 28, Czech Republic}

\begin{document}

\begin{tocentry}
\includegraphics[width=\textwidth]{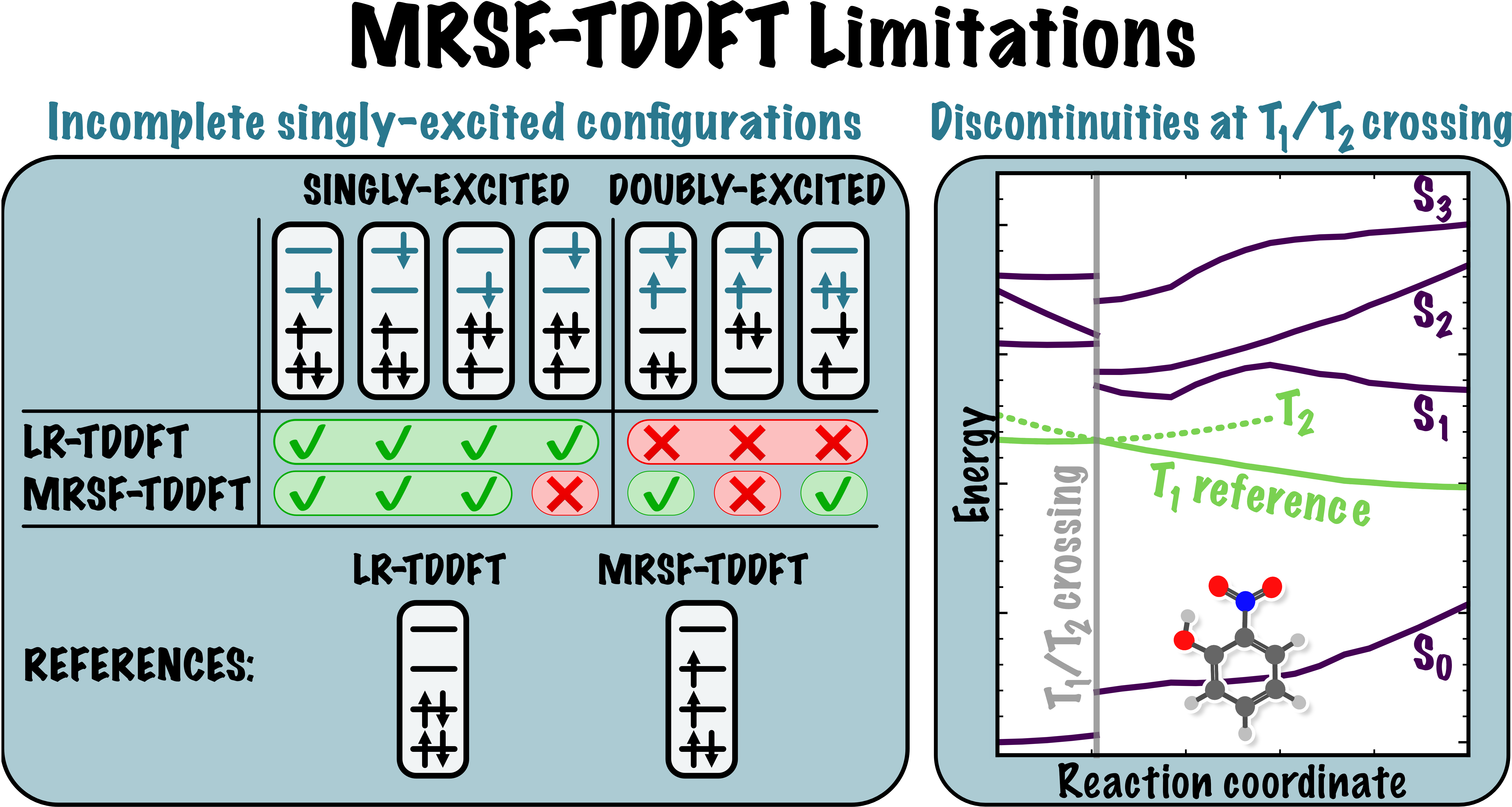}
\end{tocentry}

\renewcommand{\baselinestretch}{1.0}\normalsize

\begin{abstract}

Mixed-reference spin-flip time-dependent density functional theory (MRSF-TDDFT) has recently emerged as an attractive electronic-structure method for studying photochemical processes, given that it bridges the computational efficiency of single-reference approaches with the versatility of multireference methods. 
In the following, we critically assess the general applicability of MRSF-TDDFT to photochemistry and identify two important limitations. 
First, the doubly-excited configurations included in MRSF-TDDFT come at the cost of missing some singly-excited configurations.
Second, MRSF-TDDFT provides unreliable excited-state energies when its triplet reference---a cornerstone of the method---abruptly changes its nature, e.g., when the T$_1$ and T$_2$ triplet states become nearly degenerate and exchange electronic character. This change of character of the triplet reference can induce discontinuities or sharp distortions in electronic potential energy curves of the response states in unsuspected regions of the nuclear configuration space. 
We propose strategies and diagnostics to detect these limitations in the exploration of potential energy surfaces and nonadiabatic molecular dynamics using MRSF-TDDFT. 

\end{abstract}


\renewcommand{\baselinestretch}{1.2}\normalsize

Calculating excited-state electronic properties and simulating the dynamics of photoexcited molecules relies strongly on the quality of the underlying electronic-structure method, near the Franck-Condon region and beyond. The sensitivity of excited-state (nonadiabatic) molecular dynamics to the level of electronic-structure theory has been exemplified in multiple recent works.\cite{Plasser2014, Chakraborty2021, Janos2023, Ibele2024, Jira2024, Ibele2025, Huang2025, DeMiranda2026}  However, an efficient and easy-to-use electronic-structure method, which can reliably capture electronic energies and properties beyond the Franck--Condon region and near conical intersections, is still missing.\cite{Dreuw2026,Gonzalez2012,Knepp2025,Matsika2021,bestpractices2026}

Recently, a novel method called mixed-reference spin-flip time-dependent density functional theory (MRSF-TDDFT) has been introduced, motivated by the need for an efficient and accurate electronic-structure approach suitable for nonadiabatic dynamics.\cite{Lee2018, Lee2019} MRSF-TDDFT fills the void between single-reference and multireference methods, balancing the advantages and disadvantages of these two traditional families of electronic-structure methods and making MRSF-TDDFT appealing to computational photochemists. As a result, the development of MRSF-TDDFT has been fast-paced since its introduction in 2018, leading to a wide range of studies: the photodynamics of several molecules (cyclobutanone,\cite{Brady2025} azulene,\cite{Park2025} dihydroazulene,\cite{Shostak2023} octatetraene,\cite{Park2022a} uracil,\cite{Park2022b} cytosine,\cite{Sadiq2023} \textit{cis}-stilbene,\cite{Farmani2025} 9,9'-bifluorenylidene,\cite{Wang2026} or thymine in the gas\cite{Park2021, Komarov2023} and condensed\cite{Huix-Rotllant2023} phase), singlet-triplet gaps in diradicals,\cite{Horbatenko2021, Horbatenko2019} design of molecules for optoelectronics,\cite{Ahmad2024,Kim2021,Pradhan2020, Japahuge2019} or an atmospheric hydrogen shift reaction between peroxyradicals.\cite{Mandal2025} MRSF-TDDFT has also been implemented in several quantum-chemistry packages---OpenQP\cite{Mironov2024}, GAMESS\cite{Barca2020, Komarov2023gamess} and QChem 6.4\cite{Epifanovsky2021}--- which demonstrates the elevated interest of the theoretical chemistry community for this method. 

MRSF-TDDFT is an alternative to the well-known linear-response (LR-) TDDFT, which, in its adiabatic approximation,\footnote{The adiabatic approximation allows the use of standard ground-state exchange-correlation functionals in the context of TDDFT, yet at the cost of losing the memory effects of the dynamics of the electronic density. Implementations of LR-TDDFT in standard quantum chemistry codes employ the adiabatic approximation by default. Hence, we inherently assume the adiabatic approximation when we mention LR-TDDFT, and any methodologies emerging from TDDFT in general.} describes the excited electronic states in terms of singly-excited electronic configurations generated from the closed-shell ground-state DFT reference.\cite{tddftcarsten} LR-TDDFT is a well-established method for calculating excited-state properties in the Franck--Condon region for medium to large molecules. However, LR-TDDFT possesses several limitations: (i) the single-reference character of the underlying DFT ground electronic state, combined with approximate exchange-correlation functionals, prevents a proper description of diradicals or homolytic bond breaking,\cite{GIESBERTZ2008338, Giesbertz2008, Cordova2007} (ii) LR-TDDFT fails to describe double excitations,\cite{Maitra2004, Neugebauer2004, Levine2006} and (iii) LR-TDDFT predicts a wrong topography and topology of conical intersections involving the reference ground electronic state.\cite{Levine2006, Huix-Rotllant2013, Huix-Rotllant2015,Taylor2023, Taylor2024} Together, these features strongly restrict the application of LR-TDDFT to study photochemical processes. 

The aforementioned limitations of LR-TDDFT within the adiabatic approximation can be alleviated by introducing an open-shell triplet reference and generating a set of singlet electronic configurations via spin-flip excitations.\cite{Casanova2020} This so-called spin-flip approach possesses several advantages. First, the singlet ground electronic state becomes an 'excited state' from the perspective of the triplet reference. Therefore, it is described on an equal footing with the excited singlet electronic states, which allows a proper coupling between the excited and the ground (singlet) electronic states and leads to the correct topography and topology of conical intersections involving the ground electronic state. Moreover, constructing the ground electronic state within the response formalism allows inclusion, at least partially, of multireference character in its description. Another advantage is that the spin-flip excitations from an open-shell triplet reference can create some doubly-excited configurations\footnote{Singly-excited and doubly-excited configurations are always meant with respect to the closed-shell ground state configuration. Since the triplet reference in MRSF-TDDFT already has a singly-excited configuration, another single excitation for such a reference may create a doubly-excited configuration.} missing in LR-TDDFT due to its adiabatic approximation.
This spin-flip idea was originally developed with the $M_S=+1$ triplet reference and denoted as spin-flip TDDFT (SF-TDDFT).\cite{Shao2003, Casanova2020} However, the promising properties of SF-TDDFT have been hindered by a severe spin contamination of the response electronic states\footnote{We use the term response electronic states for electronic states generated by the response formalism of either LR-TDDFT or MRSF-TDDFT, distinguishing them from the variational reference states.} induced by the imbalanced $M_S=+1$ triplet reference.\cite{Casanova2020}

A successful solution to the spin-contamination problem in SF-TDDFT was introduced within the MRSF-TDDFT formalism by building a new reference that mixes $M_S=+1$ and $M_S=-1$ triplet states.\cite{Lee2018} By combining this mixed reference with spin-flip excitations and the Tamm--Dancoff approximation (TDA), MRSF-TDDFT effectively eliminates the spin contamination of SF-TDDFT, while keeping its advantages listed above. 
As such, MRSF-TDDFT appeared as a promising candidate in the electronic-structure market for excited electronic states, and the interested reader is referred to Refs~\citenum{Park2023, Park2025review, Lee2025} for excellent reviews about this method and the details of its formalism. Following its introduction, MRSF-TDDFT was shown to provide reliable excitation energies and singlet-triplet gaps,\cite{Horbatenko2019} yield correct double-cone topologies and topographies of conical intersections involving the ground electronic state,\cite{Lee2019ci} and describe electronic states with double-excitation character.\cite{Horbatenko2021doubleexc} Development of analytical gradients\cite{Lee2019} and nonadiabatic coupling vectors\cite{Lee2021} then allowed combination of MRSF-TDDFT with nonadiabatic molecular dynamics, leading to an ever-increasing number of applications in photochemistry, e.g., Refs.~\citenum{Brady2025,Park2025,Shostak2023,Park2022a,Park2022b,Sadiq2023,Farmani2025,Park2021, Komarov2023,Huix-Rotllant2023}. Several benchmark studies have placed MRSF-TDDFT as a suitable method for nonadiabatic molecular dynamics.\cite{Huang2025, DeMiranda2026} In these benchmarks, MRSF-TDDFT usually outperforms LR-TDDFT and sometimes even SA-CASSCF compared to (X)MS-CASPT2 references, although mainly for electronic populations (which can sometimes be a misleading or incomplete metric for the quality of a nonadiabatic molecular dynamics simulation\cite{mignolet2018walk,cigrang2025roadmapNAMD,akimov2025ranking,Ibele2025,bestpractices2026}) with no other observables having been compared. MRSF-TDDFT is also praised for its stability in contrast to electronic-structure methods based on an active space.\cite{DeMiranda2026} 

The range of applications discussed above naturally places MRSF-TDDFT as an appealing candidate for nonadiabatic molecular dynamics given its robustness, efficiency, and simplicity. The number of these applications is a testimony to the increasing interest in MRSF-TDDFT. While the fast development and capabilities of MRSF-TDDFT are highly exciting, the no-free-lunch theorem pushes us to reflect on the potential drawbacks of MRSF-TDDFT. Hence, this Letter is dedicated to an exploration of the limitations of MRSF-TDDFT for photochemical and photophysical studies, hoping that an identification of its applicability range will support the development and application of MRSF-TDDFT and prevent its improper use. Here, we formulate two limitations of the current MRSF-TDDFT formalism and demonstrate them using molecular examples.

\begin{figure}[ht!]
    \centering
    \includegraphics[width=\linewidth]{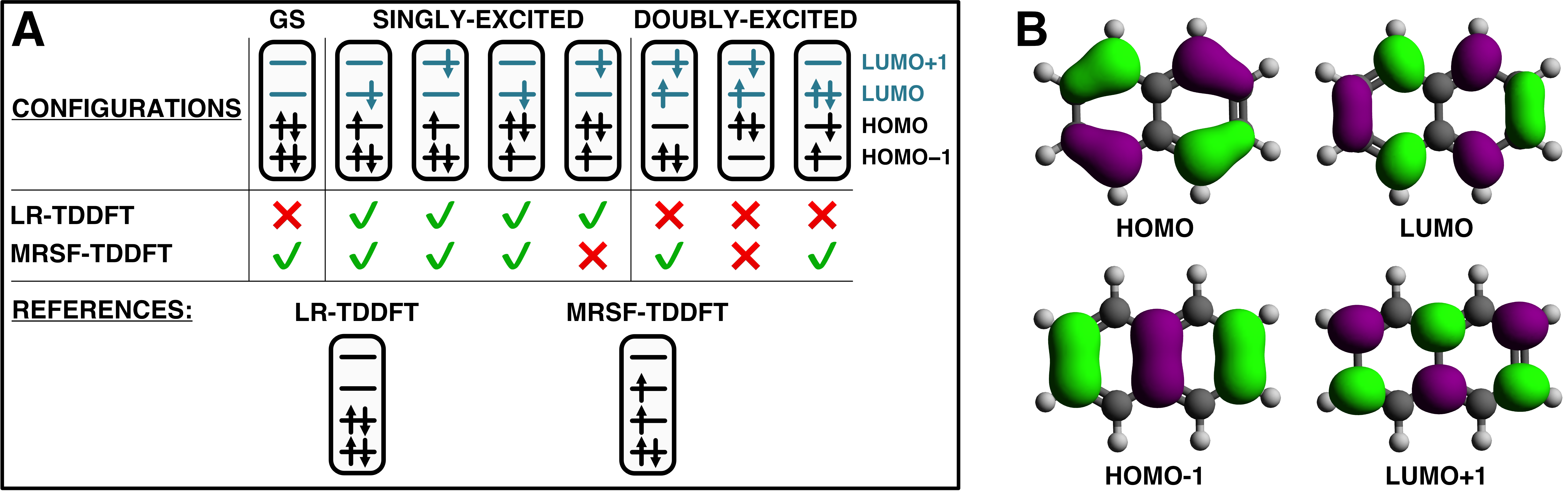}
    \caption{\textbf{A}: Schematic representation of electronic configurations which can or cannot be created in LR-TDDFT and MRSF-TDDFT from their respective references. GS stands for the ground-state configuration, which serves as the reference in LR-TDDFT, whereas this configuration is created in MRSF-TDDFT by the response formalism. The reference configuration for each method is depicted in the lower panel. For simplicity, the scheme ignores the mixed character of the reference and omits the spin balance of the excited configurations brought by MRSF-TDDFT. Configurations with electrons having opposite spin should also be included. \textbf{B}: Frontier molecular orbitals of naphthalene calculated with LRC-$\omega$PBEh/def2-SVP.}
    \label{fig:naph_orb}
\end{figure}

\textbf{Limitation 1: MRSF-TDDFT does not create a complete set of singly-excited configurations.} MRSF-TDDFT is often praised for its partial inclusion of doubly-excited configurations,\cite{Park2023, Park2025review, Lee2025,Horbatenko2021doubleexc} see Fig.~\ref{fig:naph_orb}A. However, the literature omits to mention that these doubly-excited configurations come at the cost of losing some of the singly-excited configurations present in LR-TDDFT. Specifically, any singly-excited configuration that leaves the highest occupied molecular orbital (HOMO) fully occupied and the lowest unoccupied molecular orbital (LUMO) empty will be missing in MRSF-TDDFT, because such a configuration would require a double excitation from the open-shell triplet reference used in MRSF-TDDFT. A schematic representation of configurations that can or cannot be captured by LR-TDDFT and MRSF-TDDFT is presented in Fig.~\ref{fig:naph_orb}A.

These missing singly-excited configurations may constitute a non-negligible contribution, or even be the dominant configuration, for low-lying excited electronic states. To demonstrate the impact of \textbf{Limitation 1}, we examine the excited electronic states of naphthalene, which was recently studied in Ref.~\citenum{Kimber2023}. We focus here on the four excited electronic states---$1^1B^-_{3u}$, $1^1B^+_{2u}$, $1^1B^+_{3u}$, $2^1B^+_{2u}$---that can be characterized by electronic configurations built from single excitations between the HOMO$-$1, HOMO, LUMO, and LUMO$+$1 depicted in Fig.~\ref{fig:naph_orb}B. The character of these excited electronic states can be described as follows: (\textit{i}) the $1^1B^-_{3u}$ electronic state has a mixed character of HOMO$-$1$\rightarrow$LUMO and HOMO$\rightarrow$LUMO$+$1; (\textit{ii}), the $1^1B^+_{2u}$ electronic state has HOMO$\rightarrow$LUMO character; (\textit{iii}) the $1^1B^+_{3u}$ electronic state has again a mixed character of HOMO$-$1$\rightarrow$LUMO and HOMO$\rightarrow$LUMO$+$1, just with an opposite phase; (\textit{iv}), the $2^1B^+_{2u}$ electronic state has HOMO$-$1$\rightarrow$LUMO$+$1 character. This last $2^1B^+_{2u}$ electronic state is of particular interest for our study, as it is built precisely from the type of configuration missing in MRSF-TDDFT (i.e., an electronic configuration where the HOMO remains fully occupied and the LUMO is empty). Although high in energy, the $2^1B^+_{2u}$ electronic state could be accessed experimentally by a 200~nm laser pulse, given its significant oscillator strength.

We reproduced the LR-TDDFT calculations of Ref.~\citenum{Kimber2023} employing LR-TDDFT/TDA with a LRC-$\omega$PBEh functional and a def2-SVP basis, and further validated them by using EOM-CCSD/def2-SVP, see Tab.~\ref{tab:naphtalene}. Performing an equivalent calculation with MRSF-TDDFT showed that this method indeed completely misses the $2^1B^+_{2u}$ excited electronic state with HOMO$-$1$\rightarrow$LUMO$+$1 character. This incompleteness of singly-excited configurations appears as a significant drawback of MRSF-TDDFT, which may weigh down the potential benefits of the additional doubly-excited configurations in some cases.

\begin{table}[ht!]
    \centering
    \definecolor{lightgray}{gray}{0.8}
    \definecolor{lightgray2}{gray}{0.92}
    
    \caption{Excitation energies $\Delta E$, oscillator strengths $f$, and dominant electronic configurations for selected excited electronic states of naphthalene, calculated with EOM-CCSD, LR-TDDFT/TDA, and MRSF-TDDFT. An expanded version of the table with more excited electronic states is available in SI.}
    
    \begin{threeparttable}
    \renewcommand{\arraystretch}{1.2}
    \begin{tabular}{l c c l}
        \hline
        \rowcolor{lightgray}
        State & $\Delta E$ (eV) & $f$ & Configurations\tnote{a} \\ \hline

        \rowcolor{lightgray2}
        \multicolumn{4}{c}{EOM-CCSD/def2-SVP} \\ \hline
        $1^1B^-_{3u}$ ($S_1$) & 4.51 & 0.0001 & $0.47h_{1}l - 0.46hl_1$ \\
        $1^1B^+_{2u}$ ($S_2$) & 5.37 & 0.0890 & $0.64hl$ \\
        $1^1B^+_{3u}$ ($S_4$) & 6.74 & 1.5535 & $0.47hl_1+ 0.46h_{1}l$ \\
        $2^1B^+_{2u}$ ($S_6$) & 6.93 & 0.3265 & $0.63h_{1}l_1$ \\ \hline

        \rowcolor{lightgray2}
        \multicolumn{4}{c}{LR-TDDFT/TDA/LRC-$\omega$PBEh/def2-SVP} \\ \hline
        $1^1B^-_{3u}$ ($S_1$) & 4.74 & 0.0000 & $0.70h_{1}l - 0.70hl_1$ \\
        $1^1B^+_{2u}$ ($S_2$) & 5.02 & 0.0954 & $0.93hl$ \\
        $1^1B^+_{3u}$ ($S_5$) & 6.77 & 2.0044 & $0.68hl_1+ 0.68h_{1}l$ \\
        $2^1B^+_{2u}$ ($S_6$) & 6.92 & 0.3631 & $0.89h_{1}l_1$ \\ \hline

        \rowcolor{lightgray2}
        \multicolumn{4}{c}{MRSF-TDDFT/LRC-$\omega$PBEh/def2-SVP} \\ \hline
        $1^1B^-_{3u}$ ($S_1$) & 4.34 & 0.0019 & $0.71h_{1}l + 0.67hl_1$ \\
        $1^1B^+_{2u}$ ($S_2$) & 4.63 & 0.4469 & $0.99hl$ \\
        $1^1B^+_{3u}$ ($S_3$) & 5.47 & 2.3264 & $0.72hl_1- 0.68h_{1}l$ \\
        $2^1B^+_{2u}$ (--) &\multicolumn{3}{c}{Not available}\\ \hline
    \end{tabular}

    \begin{tablenotes}\footnotesize
        \item[a] Dominant configurations and their coefficients; $h_{x}l_{y}$ refers to the HOMO$-x\rightarrow$LUMO$+y$ transition.
    \end{tablenotes}

    \end{threeparttable}
    \label{tab:naphtalene}

\end{table}

While the missing singly-excited configurations appear to hamper the applicability of MRSF-TDDFT to some molecular systems, possible extensions of MRSF-TDDFT could incorporate these missing configurations. For example, a recent preprint proposed such an extended MRSF-TDDFT strategy, but in this case to improve the description of charge-transfer states.\cite{Oh2026}
 

\textbf{Limitation 2: MRSF-TDDFT produces discontinuous or distorted potential energy surfaces in regions of nuclear configuration space where the underlying triplet reference changes electronic character, particularly when $T_1$ and $T_2$ become (nearly) degenerate.} One of the main deficiencies of LR-TDDFT appears when the $S_0$ and $S_1$ electronic states become degenerate, resulting in wrong topologies and topographies of the resulting conical intersections\cite{Levine2006, Huix-Rotllant2013, Huix-Rotllant2015,Taylor2023, Taylor2024} and, often, negative excitation energies. The problem is two-fold: first, the degenerate ground-state density is poorly described by DFT (with approximate exchange-correlation functionals), and second, the coupling between the variational ground-state reference and the response first excited electronic state is missing within the adiabatic approximation. Although MRSF-TDDFT fixes the issue of the wrong topology/topography of conical intersection by describing both the ground and excited electronic states within the response formalism,\cite{Lee2019ci} the method is not free of the problem emerging with a degenerate reference. In the case of MRSF-TDDFT, it is the $T_1$/$T_2$ degeneracy that causes problems for the reference, instead of the $S_0$/$S_1$ degeneracy for LR-TDDFT. To the best of our knowledge, this limitation of MRSF-TDDFT has remained unnoticed to date in the literature.\cite{Park2023, Park2025review, Lee2025}

\begin{figure}[ht!]
    \centering
    \includegraphics[width=1\linewidth]{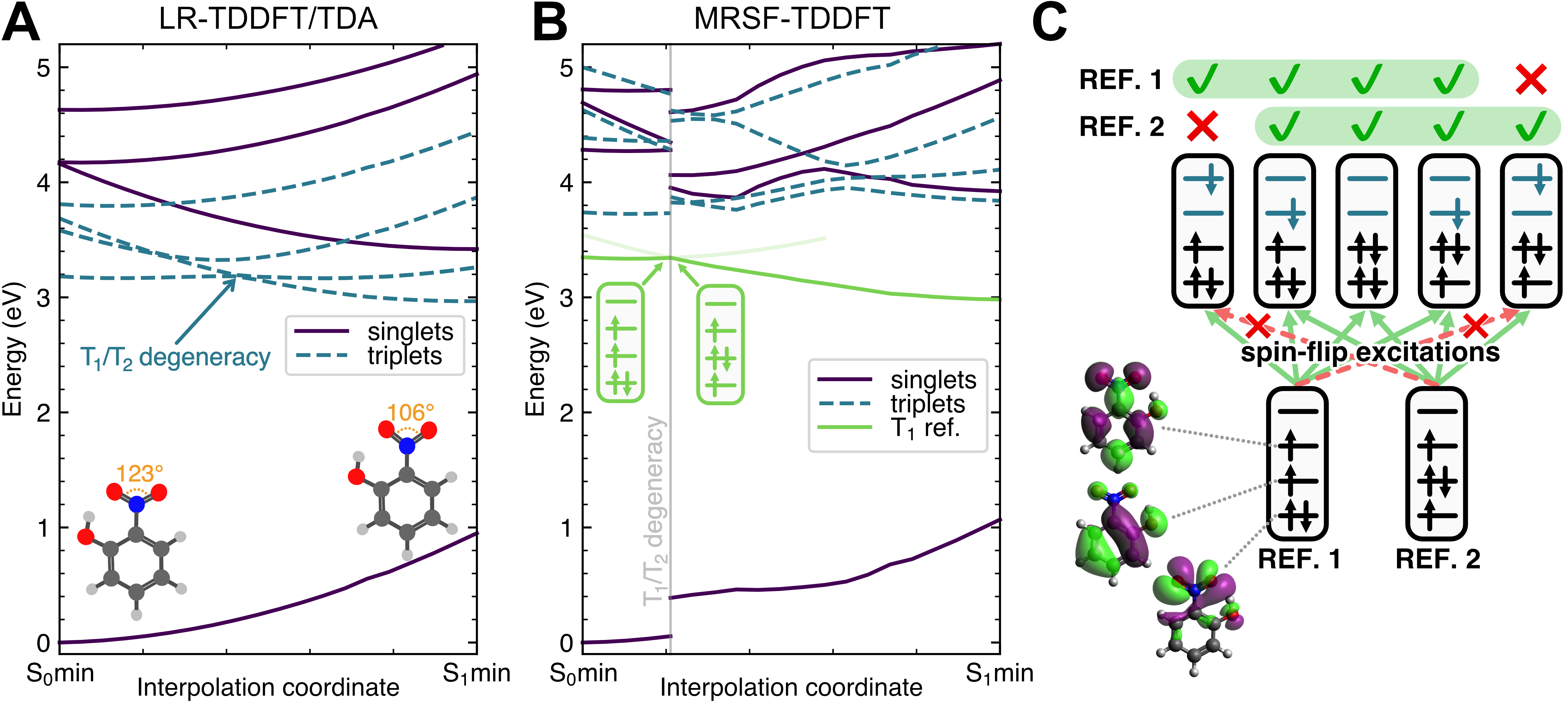}
    \caption{Electronic energies along a geodesic interpolation between minima on the $S_0$ and $S_1$ electronic states of \textit{ortho}-nitrophenol. The two geometries, taken from Ref.~\citenum{Greene2024}, differ by a small alteration of the O--N--O angle. \textbf{A}: LR-TDDFT/TDA/$\omega$B97X-D/def2-TZVP calculations. \textbf{B}: MRSF-TDDFT/$\omega$B97X-D/def2-TZVP calculations. The plot contains the reference triplet state (solid green line) and the response triplet states (dashed blue lines). The $T_1$/$T_2$ point of degeneracy for the reference triplet state is marked by a grey vertical line. The electronic configurations of the reference at each side of the degeneracy are depicted as an inset. \textbf{C}: Schematic depiction of configurations created by spin-flip excitations in MRSF-TDDFT from the two reference triplet configurations. An expanded scheme with more configurations can be found in the SI.}
    \label{fig:onp_pes}
\end{figure}

To demonstrate \textbf{Limitation 2}, we introduce two molecular systems---\textit{ortho}-nitrophenol and ethyl diazoacetate---each suffering from triplet-reference instabilities at $T_1$/$T_2$ crossings, although the two are affected differently. Let us start with \textit{ortho}-nitrophenol, where we focus on the previously studied initial nonradiative relaxation of the molecule excited to its $S_1$ state.\cite{Greene2024} The LR-TDDFT potential energy curves in Fig.~\ref{fig:onp_pes}A (produced using a geodesic interpolation\cite{Zhu2019}) show that \textit{ortho}-nitrophenol excited to $S_1$ can directly relax to the $S_1$ minimum without encountering any barrier. However, the $T_1$ and $T_2$ electronic states become degenerate along this relaxation pathway. While this degeneracy is irrelevant for the singlet nonradiative decay of \textit{ortho}-nitrophenol and for the LR-TDDFT calculations in general, it plays a crucial role for MRSF-TDDFT.

Potential energy curves calculated with MRSF-TDDFT present a different picture (Fig.~\ref{fig:onp_pes}B). The pathway is divided into two regions by the $T_1$/$T_2$ point of degeneracy, where the reference triplet state changes its electronic character. Namely, one electron from the highest doubly-occupied orbital moves to the lowest singly occupied orbital. Hence, both references still involve the same occupied molecular orbitals, but their occupation now differs. The $T_1$/$T_2$ degeneracy is a sharp local feature along the interpolation path, leading to an abrupt change of character for the triplet reference. Since MRSF-TDDFT creates only single spin-flip excitations, the sets of electronic configurations generated from the two references (before and after the point of degeneracy) are not the same, although they overlap significantly (Fig.~\ref{fig:onp_pes}C). Because the two sets of configurations differ, MRSF-TDDFT does not create the same response electronic states before and after the $T_1$/$T_2$ point of degeneracy. Thus, the response MRSF-TDDFT electronic states (singlets and triplets) do not connect at the interface of the two regions, creating a discontinuity in all the potential energy curves (Fig.~\ref{fig:onp_pes}B). Moreover, the electronic states do not even keep the same character/slope in the two regions, which is best seen with the triplet states (Fig.~\ref{fig:onp_pes}B). 
Therefore, nonadiabatic molecular dynamics simulations of nonradiative relaxation of \textit{ortho}-nitrophenol (photoexcited to its $S_1$ electronic state) would suffer from severe energy discontinuities with MRSF-TDDFT. Similar energy discontinuities appear in multiconfigurational/multireference methods when orbitals in the active space change and are a reason for discarding such trajectories (see Ref.~\citenum{bestpractices2026} for a discussion and an example of such an issue).

\begin{figure}[ht!]
    \centering
    \includegraphics[width=1.0\linewidth]{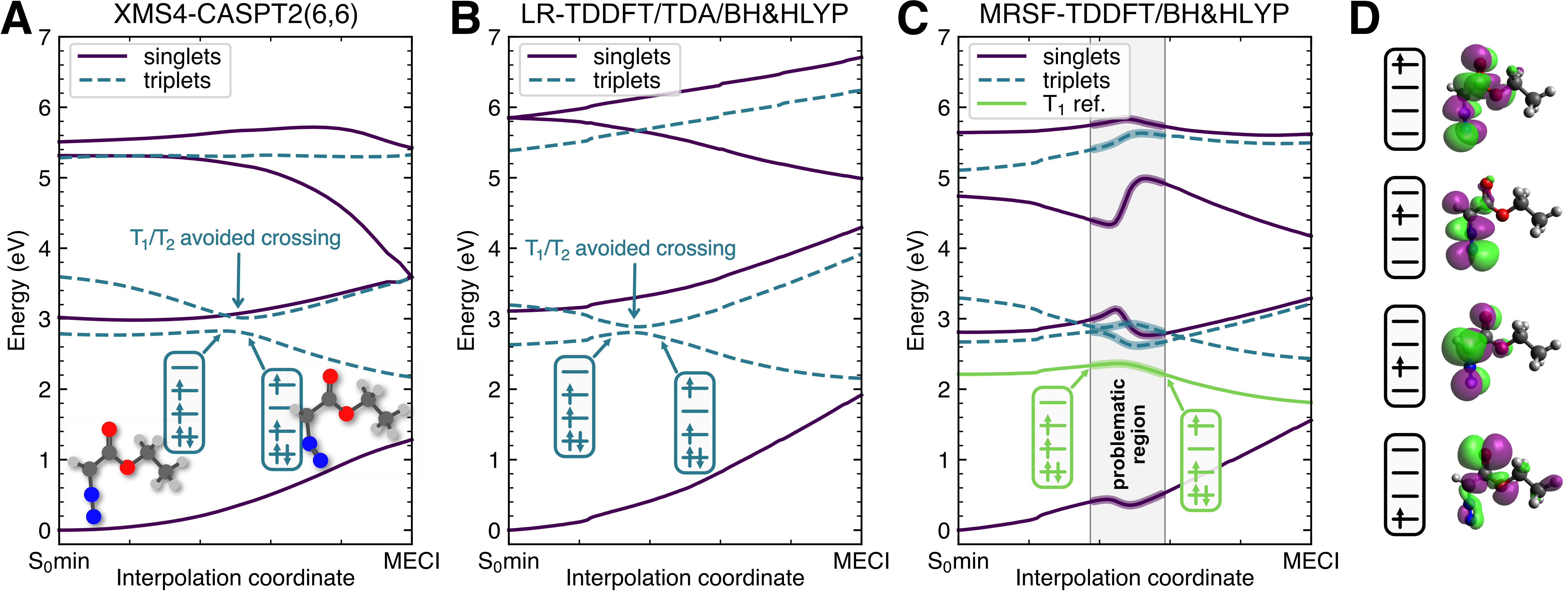}
    \caption{Electronic energies along a geodesic interpolation between the $S_0$ minimum and the $S_2$/$S_1$ MECI of ethyl diazoacetate, optimized at the XMS4-CASPT2(6,6)/cc-pVDZ level of theory. \textbf{A}: XMS-CASPT2(6,6)/cc-pVDZ calculation. \textbf{B}: LR-TDDFT/TDA/BH\&HLYP/cc-pVDZ calculation. \textbf{C}: MRSF-TDDFT/BH\&HLYP/cc-pVDZ calculation. The plot contains the reference triplet state (solid green line) and the response triplet states (dashed blue lines) calculated with MRSF-TDDFT. The problematic region around the $T_1$/$T_2$ avoided crossing of the reference triplet state is marked by a grey rectangle. \textbf{D}: Four frontier molecular orbitals of ethyl diazoacetate.}
    \label{fig:eda_pes}
\end{figure}

Let us now move to the photodynamics of the second molecule, ethyl diazoacetate, to demonstrate a different aspect of \textbf{Limitation 2}. Ethyl diazoacetate can undergo a Wolff rearrangement upon photoexcitation to its $S_2$ electronic state by a 270-nm laser pulse.\cite{Phelps2022} Here, we focus on the initial step of the reaction by calculating potential energy curves between the $S_0$ minimum and the minimum energy conical intersection (MECI) between the $S_2$ and $S_1$ electronic states. We use XMS-CASPT2 as a reference for the electronic energies, showing a barrierless route from the Franck-Condon region ($S_0$ minimum) towards the MECI (Fig.~\ref{fig:eda_pes}A). Monitoring the $T_1$ and $T_2$ electronic states of ethyl diazoacetate along this interpolation pathway reveals that they undergo an avoided crossing, even though this observation is irrelevant to the deactivation process of the molecule in its $S_2$ electronic state. LR-TDDFT shows qualitatively comparable results to XMS-CASPT2, except for the MECI geometry (optimized at the XMS-CASPT2 level of theory), where the energy gap is about 1 eV, see Fig.~\ref{fig:eda_pes}B. LR-TDDFT would most likely predict a slightly different geometry for the MECI, yet it is not the scope of this work to localize this structure. Irrespective of the MECI position, LR-TDDFT exhibits smooth curves, which would allow stable nonadiabatic molecular dynamics simulations of this nonradiative process. 

Contrary to LR-TDDFT, MRSF-TDDFT provides pathologic potential energy curves along the interpolation pathway due to the underlying $T_1$/$T_2$ avoided crossing (Fig.~\ref{fig:eda_pes}C). The $T_1$/$T_2$ avoided crossing again splits the interpolation pathway into two regions with a different electronic configuration for the triplet reference and incompatible response electronic states, as we observed earlier for the \textit{ortho}-nitrophenol case. 
Despite this apparent similarity, we stress here two key differences between the ethyl diazoacetate and \textit{ortho}-nitrophenol cases: (\textit{i}) the $T_1$/$T_2$ avoided crossing is not abrupt/localized and the triplet states smoothly change character over a certain region of configuration space (Fig.~\ref{fig:eda_pes}C); (\textit{ii}) the two resulting triplet references do not occupy the same set of molecular orbitals this time (Fig.~\ref{fig:eda_pes}D), but one of the unpaired electrons changes its orbital between the two references (all the doubly occupied orbitals remain the same). As a consequence, the MRSF-TDDFT potential energy curves for ethyl diazoacetate exhibit a \textquote{problematic region} instead of the sharp discontinuity seen for the electronic energies of \textit{ortho}-nitrophenol. In this \textquote{problematic region}, the potential energy curves distort dramatically to connect the electronic states on the two sides of the $T_1$/$T_2$ avoided crossing. We note that the result is independent of the functional, as demonstrated with CAM-B3LYP in the SI.

Hence, \textbf{Limitation 2} always manifests itself as a result of two incompatible sets of response states, stemming from different triplet reference configurations on each side (region) of a $T_1$/$T_2$ avoided crossing or point of degeneracy. The way the potential energy curves behave at the interface of these two regions then depends on the nature of the $T_1$/$T_2$ interaction. For a $T_1$/$T_2$ point of degeneracy, i.e., an abrupt change of character, a sharp discontinuity is observed in the resulting MRSF-TDDFT potential energy curves. For an avoided crossing between $T_1$ and $T_2$, i.e., the character of the triplet state varies smoothly, the potential energy curves are distorted yet continuous.
To confirm that the implementation of MRSF-TDDFT was not the source of the problems observed here, we compared the results of MRSF-TDDFT obtained with the OpenQP\cite{Mironov2024} and GAMESS\cite{Barca2020} codes; no differences were noticed.

In any case, we foresee serious issues if MRSF-TDDFT is used for nonadiabatic molecular dynamics when a molecular system explores regions of configuration space where the underlying $T_1$ and $T_2$ electronic states come close in energy (independently of whether the dynamics takes place in the singlet or triplet manifold). While the first case of sharp discontinuities ($T_1$/$T_2$ degeneracy) can, in principle, be easily spotted in the dynamics, the second case ($T_1$/$T_2$ avoided crossing) results in distorted potential energy curves that might be harder to detect. Such discontinuities and distortions, possibly related to \textbf{Limitation 2}, can already be spotted in published works using MRSF-TDDFT for nonadiabatic molecular dynamics.\cite{Huang2025, Brady2025, Sadiq2023}


Thus, MRSF-TDDFT in its current formulation should not be used for dynamics that encounter regions where the $T_1$ and $T_2$ states come close in energy, somewhat echoing the fact that LR-TDDFT should be avoided for systems exhibiting $S_0$/$S_1$ conical intersections. However, the risk for MRSF-TDDFT appears higher than for LR-TDDFT, because $S_0$ and $S_1$ are typically calculated along a trajectory using LR-TDDFT and it is simple to detect when they come close in energy. On the contrary, the triplet states are not always monitored when the dynamics are dominated by singlet nonradiative channels, and a degeneracy (or near-degeneracy) of the $T_1$ and $T_2$ electronic states might pass unnoticed, especially if it manifests itself on the MRSF-TDDFT energies only as a continuous distortion of the potential energy curves and not an abrupt discontinuity.

To alleviate the danger of misusing MRSF-TDDFT in nonadiabatic molecular dynamics, we suggest here some practical diagnostics that a user of MRSF-TDDFT can use to detect the consequences of \textbf{Limitation 2} highlighted in this work. The most efficient diagnostic, even if computationally demanding, is to calculate the response triplet electronic states during the nonadiabatic molecular dynamics and to monitor the energy gap between the $T_1$ and $T_2$ electronic states. This energy gap is the best marker of a possible problem with the reference, and making sure it remains larger than a certain threshold (for example, 0.1-0.2 eV) should prevent an issue. A less computationally-demanding diagnostic consists of monitoring the overlaps of the singly occupied orbitals between different time steps to detect a possible change in the character of the triplet reference. Finally, the simplest (yet least reliable) diagnostics we can suggest is to monitor the energy gap between the two singly-occupied orbitals in an MRSF-TDDFT calculation, as they could become degenerate when $T_1$ and $T_2$ come close in energy. 

Ongoing developments of MRSF-TDDFT are likely to improve upon the limitations flagged in this work. First, extending the number of electronic configurations generated by MRSF-TDDFT to include the missing singly-excited configurations (\textbf{Limitation 1}) could make the sets of electronic configurations generated from two different triplet references more compatible, leading to less severe discontinuities or distortions (\textbf{Limitation 2}).\cite{Oh2026} Nevertheless, different sets of doubly-excited configurations will still keep the two sets of electronic configurations incompatible (see SI for a schematic representation). Second, applying an unrestricted DFT formalism could bring additional flexibility to the triplet reference and improve the stability of the calculation over the current restricted open-shell version.\cite{Komarov2024, Makhnev2026} Problems with the convergence of the restricted open-shell triplet reference are known in the literature,\cite{Casanova2020, Brady2025} and also appear in our calculations in the vicinity of the $T_1$/$T_2$ (near-) degeneracy.

In summary, we have demonstrated two limitations of the current MRSF-TDDFT formalism that remained unnoticed in the literature and restrict its application in photochemistry. The first limitation is the incomplete set of singly-excited configurations produced by MRSF-TDDFT, which comes at the cost of having some additional doubly-excited configurations. The missing singly-excited configurations represent excitations where the HOMO remains fully occupied and the LUMO is empty. These configurations might constitute partial or even dominant contributions to excited states, as illustrated by the excited electronic states of naphthalene. The second limitation stems from the underlying triplet reference, which may change character in a part of nuclear configuration space where the underlying $T_1$ and $T_2$ electronic states come close in energy. In the surrounding region of configuration space, the MRSF-TDDFT response electronic states exhibit either sharp discontinuities or sudden distortions, illustrated here by the potential energy curves of \textit{ortho}-nitrophenol and ethyl diazoacetate. Nonadiabatic molecular dynamics simulations passing through the vicinity of a $T_1$/$T_2$ (near-) degeneracy would be influenced by these discontinuities or distortions, even if they follow a singlet electronic state and no dynamics are considered in the triplet manifold. 
We remark that the above mentioned limitations would have an impact on any spin-flip approaches based on single excitations from a triplet reference, such as standard SF-TDDFT or spin-adapted SF-TDDFT.\cite{Casanova2020, Zhang2015}
Finally, we suggested strategies to diagnose this second limitation in nonadiabatic molecular dynamics simulations, as its consequences can be pernicious. As a result, MRSF-TDDFT remains a powerful and reliable electronic-structure method for photochemistry, if combined with proper benchmarking and on-the-fly diagnostics. 

\section*{Computational Details}

For naphthalene, the following codes were used: Gaussian 16, Revision A.03\cite{g16} for EOM-CCSD/def2-SVP; QChem 6.2\cite{Epifanovsky2021} for LR-TDDFT/TDA/LRC-$\omega$PBEh/def2-SVP with the input file provided in Ref.~\citenum{Kimber2023} (only with six excited states instead of four); OpenQP version 1.0 Aug, 2024\cite{Mironov2024} for MRSF-TDDFT/LRC-$\omega$PBEh/def2-SVP.

In the case of \textit{ortho}-nitrophenol, LR-TDDFT/TDA/$\omega$B97X-D/def2-TZVP was performed in Gaussian 16 Revision A.03,\cite{g16} while MRSF-TDDFT/$\omega$B97X-D/def2-TZVP was done in GAMESS 2024 R2.\cite{Barca2020, Komarov2023gamess} We used GAMESS instead of OpenQP in this case as the reference ROKS calculation was more stable in GAMESS. In the regions of nuclear configuration space where both codes converged to the same reference ROKS solution, the response MRSF-TDDFT states provided by the two codes were the same within numerical error. The geodesic interpolation used for Fig.~\ref{fig:onp_pes} was generated with a code provided in Ref.~\citenum{Zhu2019}. The $S_0$ and $S_1$ minimum geometries were taken from the Supporting information of Ref.~\citenum{Greene2024}.

The quantum chemistry codes used for EDA were as follows: XMS4-CASPT2(6,6)/cc-pVDZ was performed in BAGEL,\cite{Shiozaki2018} LR-TDDFT/TDA/BH\&HLYP/cc-pVDZ in Gaussian 16 Revision A.03,\cite{g16} and MRSF-TDDFT/BH\&HLYP/cc-pVDZ in OpenQP version 1.0 Aug 2024.\cite{Mironov2024} The XMS4-CASPT2(6,6) calculation was averaged over four excited electronic states with an active space of six electrons in six orbitals (see SI for the orbitals). The geodesic interpolation was again generated with the code provided in Ref.~\citenum{Zhu2019}. The interpolated geometries---$S_0$ minimum and minimum energy conical intersection between $S_2$ and $S_1$---were optimized at the XMS4-CASPT2(6,6) level of theory.

\begin{acknowledgement}

JJ and PS thank the Czech Science Foundation for support via grant number 26-22810S.
This work was supported by the project "The Energy Conversion and Storage", funded as project No. CZ.02.01.01/00/22\textunderscore 008/0004617 by Programme Johannes Amos Comenius, call Excellent Research, and by the grant of Specific university research – grant No A2\_FCHI\_2026\_051.
BFEC acknowledges funding from EPSRC for the grants EP/V026690/1, EP/Y01930X/1, and EP/X026973/1.
AJOE acknowledges funding from EPSRC for the grant EP/V026690/1, ERC Advanced Grant 101199104 and Leverhulme Research Fellowship RF-2025-194$\backslash$4.

\end{acknowledgement}

\begin{suppinfo}

Full characterization of the excited electronic states of naphthalene, analysis of the electronic configurations generated by MRSF-TDDFT with different triplet references, ethyl diazoacetate results with the CAM-B3LYP functional, active-space orbitals used in XMS4-CASPT2(6,6). (PDF) \\
All the input and output files of the calculations presented in this work are available on Zenodo (link to be included upon acceptance).

\end{suppinfo}


\providecommand{\latin}[1]{#1}
\makeatletter
\providecommand{\doi}
  {\begingroup\let\do\@makeother\dospecials
  \catcode`\{=1 \catcode`\}=2 \doi@aux}
\providecommand{\doi@aux}[1]{\endgroup\texttt{#1}}
\makeatother
\providecommand*\mcitethebibliography{\thebibliography}
\csname @ifundefined\endcsname{endmcitethebibliography}  {\let\endmcitethebibliography\endthebibliography}{}

\end{document}


\maketitle
\singlespacing
{ \hypersetup{hidelinks} \tableofcontents }

\clearpage
\section{Full characterization of the excited electronic states of naphthalene}

While Ref.~\citenum{Kimber2023} reports the $1^1B^-_{3u}$, $1^1B^+_{2u}$, $1^1B^+_{3u}$ and $2^1B^+_{2u}$ states as the lowest four adiabatic excited states $S_1$--$S_4$, their calculation was not converged with respect to the number of excited states. We converged the calculation in that respect and showed that two states were missing, in agreement with Ref.~\citenum{Hashimoto1996}, see Tab.~\ref{tab:naphtalene}. One of these states was also not captured by MRSF-TDDFT as it exhibits a mixed character of HOMO$-$1$\rightarrow$LUMO$+$2 and HOMO$-$2$\rightarrow$LUMO$+$1 transitions, further emphasizing the impact of \textbf{Limitation 1}.

\begin{table}[ht!]
    \centering
    \definecolor{lightgray}{gray}{0.8}
    \definecolor{lightgray2}{gray}{0.92}
    \begin{threeparttable}
    \renewcommand{\arraystretch}{1.2}
    \begin{tabular}{l c c l}
        \hline
        \rowcolor{lightgray}
        State\tnote{a} & $\Delta E$ (eV) & $f$ & Configurations\tnote{a} \\ \hline

        \rowcolor{lightgray2}
        \multicolumn{4}{c}{EOM-CCSD/def2-SVP} \\ \hline
        $1^1B^-_{3u}$ ($S_1$) & 4.51 & 0.0001 & $0.47h_{1}l - 0.46hl_1$ \\
        $1^1B^+_{2u}$ ($S_2$) & 5.37 & 0.0890 & $0.64hl$ \\        
        $2^1A^+_{g}$ ($S_3$) & 6.35 & 0.0000 & $0.45h_{1}l_2 + 0.44h_{2}l_1$ \\
        $1^1B^+_{3u}$ ($S_4$) & 6.74 & 1.5535 & $0.47hl_1+ 0.46h_{1}l$ \\
        $1^1B^+_{1g}$ ($S_5$) & 6.82 & 0.0000 & $0.51hl_2 + 0.41h_{2}l$ \\
        $2^1B^+_{2u}$ ($S_6$) & 6.93 & 0.3265 & $0.63h_{1}l_1$ \\ \hline

        \rowcolor{lightgray2}
        \multicolumn{4}{c}{TDA/TDDFT/LRC-$\omega$PBEh/def2-SVP} \\ \hline
        $1^1B^-_{3u}$ ($S_1$) & 4.74 & 0.0000 & $0.70h_{1}l - 0.70hl_1$ \\
        $1^1B^+_{2u}$ ($S_2$) & 5.02 & 0.0954 & $0.93hl$ \\
        $1^1B^-_{1g}$ ($S_3$) & 6.52 & 0.0000 & $0.77hl_2 - 0.63h_{2}l$ \\
        $2^1A^-_{g}$ ($S_4$) & 6.61 & 0.0000 & $0.68h_{1}l_2 - 0.66h_{2}l_1$ \\
        $1^1B^+_{3u}$ ($S_5$) & 6.77 & 2.0044 & $0.68hl_1+ 0.68h_{1}l$ \\
        $2^1B^+_{2u}$ ($S_6$) & 6.92 & 0.3631 & $0.89h_{1}l_1$ \\ \hline

        \rowcolor{lightgray2}
        \multicolumn{4}{c}{MRSF-TDDFT/LRC-$\omega$PBEh/def2-SVP} \\ \hline
        $1^1B^-_{3u}$ ($S_1$) & 4.34 & 0.0019 & $0.71h_{1}l + 0.67hl_1$ \\
        $1^1B^+_{2u}$ ($S_2$) & 4.63 & 0.4469 & $0.99hl$ \\
        $1^1B^+_{3u}$ ($S_3$) & 5.47 & 2.3264 & $0.72hl_1- 0.68h_{1}l$ \\
        $1^1B^-_{1g}$ ($S_4$) & 5.89 & 0.0000 & $0.68hl_2 - 0.69h_{2}l$ \\
        $1^1B^+_{1g}$ ($S_5$) & 6.18 & 0.0000 & $0.71hl_2 + 0.69h_{2}l$ \\  
        $2^1B^+_{2u}$ (--) &\multicolumn{3}{c}{Not available}\\ \hline
    \end{tabular}

    \begin{tablenotes}\footnotesize
        \item[a] Dominant configurations and coefficients; $h_{x}l_{y}$ refers to the HOMO$-x\rightarrow$LUMO$+y$ transition.
    \end{tablenotes}
    \end{threeparttable}

    \caption{Excitation energies $\Delta E$, oscillator strengths $f$, and dominant electronic configurations for selected excited electronic states of naphthalene calculated with different electronic-structure methods. The terms of the electronic states were identified from Ref.~\citenum{Hashimoto1996}.}

    \label{tab:naphtalene}
\end{table}

\clearpage
\section{Analysis of the electronic configurations generated by MRSF-TDDFT with different triplet references}

\begin{figure}[ht!]
    \centering
    \includegraphics[width=0.8\linewidth]{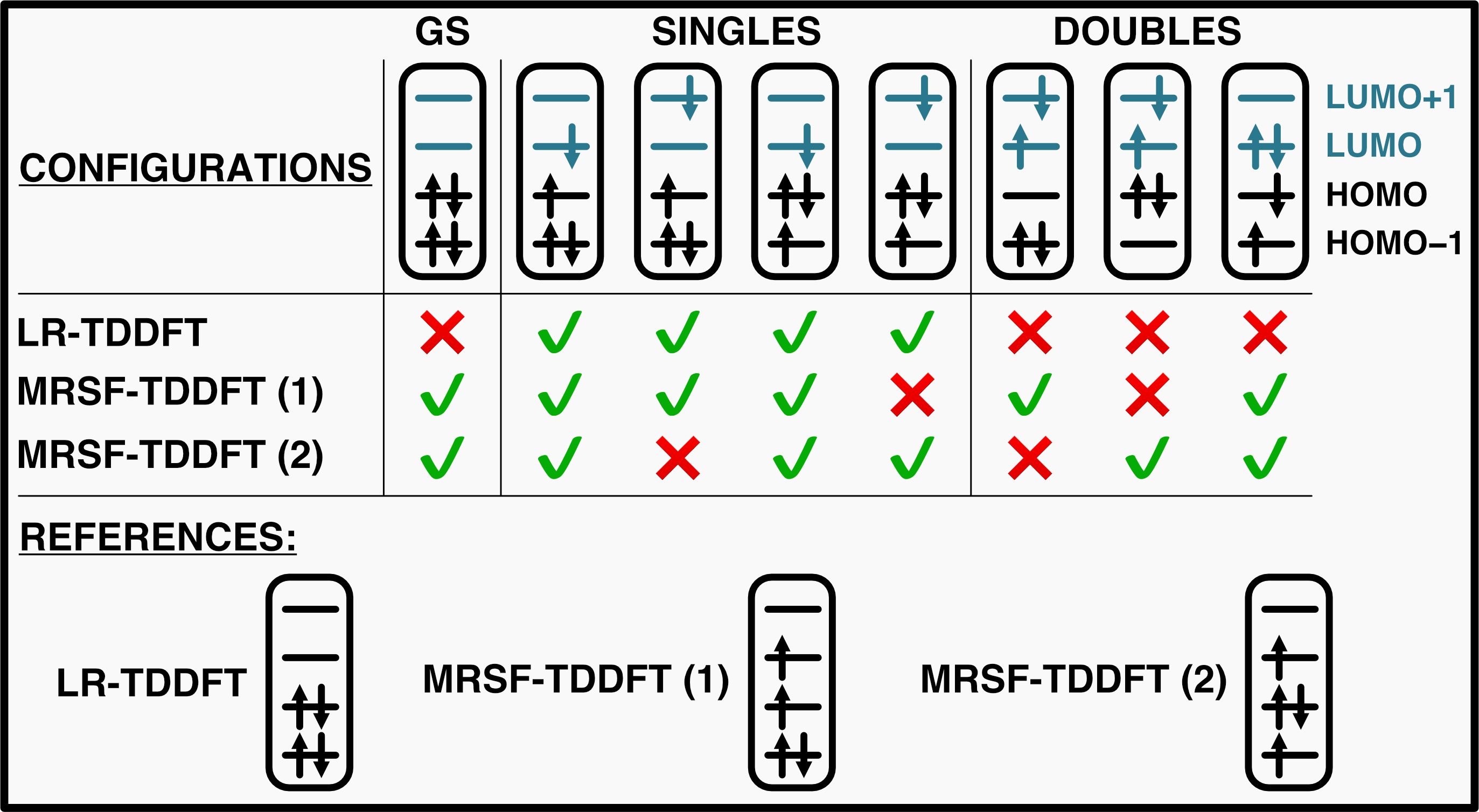}
    \caption{Simplified schematic of the electronic configurations generated by LR-TDDFT and MRSF-TDDFT from two references with different singly-occupied orbitals, as appearing in the case of \textit{ortho}-nitrophenol. The schematic demonstrates that having different occupations of orbitals in the triplet reference of MRSF-TDDFT leads to not fully-overlapping sets of electronic configurations, which are then used to generate the response electronic states. Hence, the response electronic states generated from the two triplet references would not be the same in general. Note that extending the MRSF-TDDFT formalism to include the missing singly-excited configurations would enlarge the overlap between the sets, yet the doubly-excited configurations would still make a difference.}
    \label{fig:configs}
\end{figure}

\clearpage
\section{Ethyl diazoacetate results with the CAM-B3LYP functional}

\begin{figure}[ht!]
    \centering
    \includegraphics[width=\linewidth]{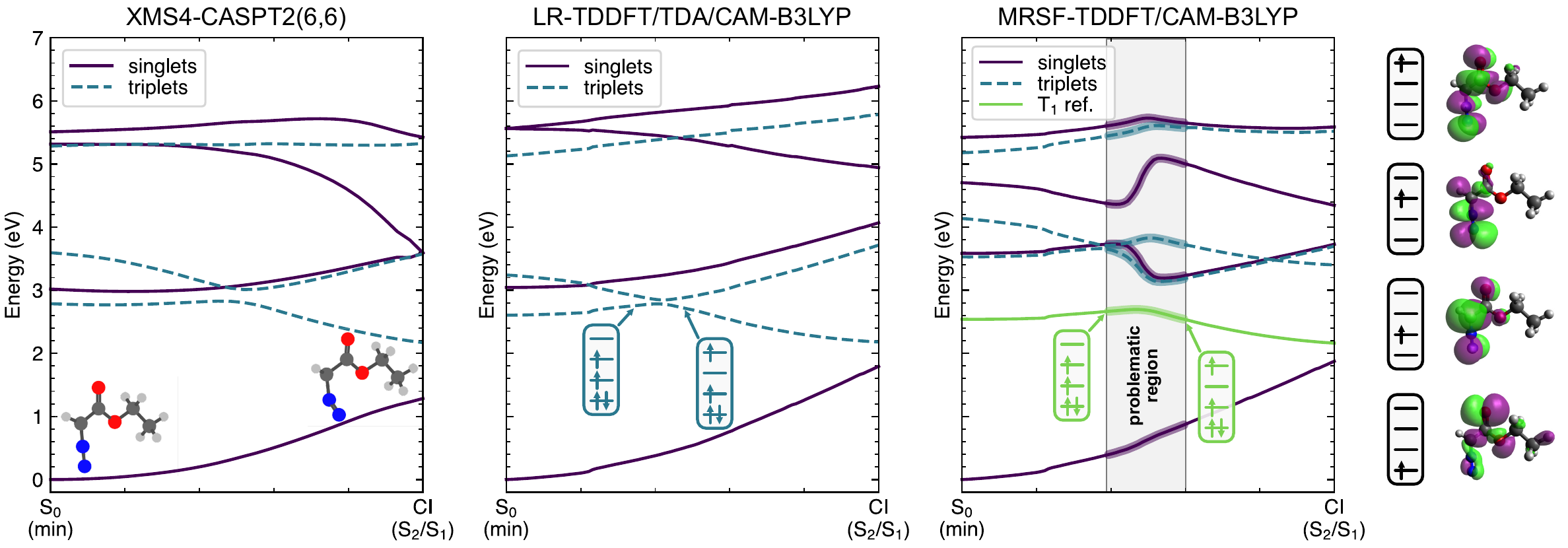}
    \caption{Electronic energies along a geodesic interpolation between the $S_0$ minimum and the $S_2$/$S_1$ MECI of ethyl diazoacetate, optimized at XMS4-CASPT2(6,6)/cc-pVDZ level of theory. \textbf{A}: XMS-CASPT2(6,6)/cc-pVDZ calculation. \textbf{B}: LR-TDDFT/TDA/CAM-B3lYP/cc-pVDZ calculation. \textbf{C}: MRSF-TDDFT/CAM-B3lYP/cc-pVDZ calculation. The plot contains the reference triplet state (solid green line) and the response triplet states (dashed blue lines) calculated with MRSF-TDDFT. The problematic region around the $T_1$/$T_2$ avoided crossing of the reference triplet state is marked by a grey rectangle. \textbf{D} Four frontier orbitals of ethyl diazoacetate.}
    \label{fig:eda}
\end{figure}

\clearpage
\section{Active-space orbitals used in XMS4-CASPT2(6,6)}

\begin{figure}[ht!]
    \centering
    \includegraphics[width=0.8\linewidth]{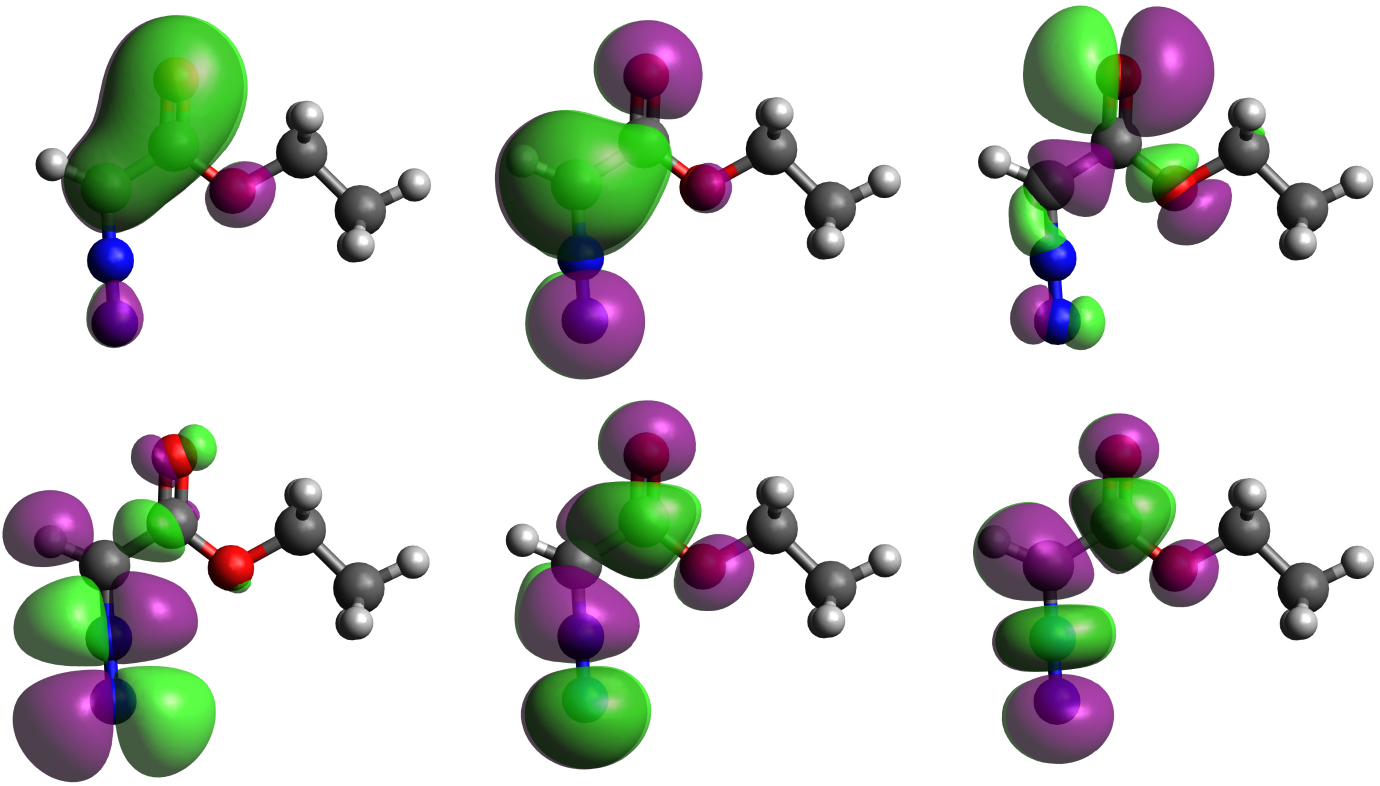}
    \caption{Orbitals composing the active space of the XMS4-CASPT2(6,6)/cc-pVDZ method used for ethyl diazoacetate.}
    \label{fig:eda_as}
\end{figure}

\clearpage

\providecommand{\latin}[1]{#1}
\makeatletter
\providecommand{\doi}
  {\begingroup\let\do\@makeother\dospecials
  \catcode`\{=1 \catcode`\}=2 \doi@aux}
\providecommand{\doi@aux}[1]{\endgroup\texttt{#1}}
\makeatother
\providecommand*\mcitethebibliography{\thebibliography}
\csname @ifundefined\endcsname{endmcitethebibliography}  {\let\endmcitethebibliography\endthebibliography}{}